\documentstyle[psfig]{article}
\begin{document}
\title{Level density and $\gamma$ strength function in $^{162}$Dy from 
inelastic $^3$He scattering}
\author{A. Schiller, M. Guttormsen, E. Melby, J. Rekstad, and S. Siem\\
Department of Physics, University of Oslo,\\ 
P.O.Box 1048, Blindern, N-0316 Oslo, Norway}
\date{\today}
\maketitle
\begin{abstract}
Complementary measurements have been performed for the level density and 
$\gamma$ strength function in $^{162}$Dy using inelastic $^3$He scattering. 
Comparing these results to previous measurements using the 
$^{163}$Dy($^3$He,$\alpha$) reaction, reveals that the measured quantities 
above 1.5~MeV do not depend significantly on the nuclear reaction chosen.

\noindent
PACS number(s): 21.10.Ma, 21.10.Pc, 25.55.-e, 27.70.+q
\end{abstract}

\section{Introduction}

Nuclear level densities have recently gained new interest. When earlier studies
of level densities were mainly based on counting levels close to the ground 
state and neutron resonance spacing at the neutron binding energy 
\cite{GC65,ES88}, a variety of new methods and experimental results are 
available today. A more recent compilation of all existing data on level 
densities \cite{IM92} includes level spacing data of several other reactions
involving light particles up to $A=4$ as well as results from Ericson 
fluctuation measurements. Recently, experimental level densities in $^{69}$As 
and $^{70}$Ge over a large excitation energy interval of 5-24~MeV have been 
reported \cite{PC99}, obtained from proton evaporation spectra of $^{12}$C 
induced reactions. Also the Oslo cyclotron group has reported on a new method 
to extract level density and $\gamma$ strength function from primary $\gamma$
spectra (see \cite{HB95} for the basic assumptions and \cite{SB99} for the 
method). This method has the advantage that the level density is deduced from 
$\gamma$ transitions, thus the nucleus is likely to be thermalized and the 
measured level density is supposed to be independent of the formation mechanism
of the excited nucleus. Several applications of the method are reported in
\cite{MB99,SB99b,GH99,GB99}.

The experimental progress has been accompanied by new theoretical developments.
with respect to the first analytical nuclear level density formula proposed by 
Bethe \cite{Be36}. Level densities have been studied for finite temperatures 
within the BCS model \cite{SY63,Ng90}. Today, Monte Carlo shell model 
calculations \cite{LJ93,KD97} are able to estimate nuclear level densities 
\cite{Or97} for heavy mid shell nuclei like $^{162}$Dy \cite{WK98}. Also more 
schematic approaches like binomial level densities \cite{Zu99} have been 
revived lately. Important applications of the theoretical and experimental 
efforts are calculations of the nucleon synthesis in stars, where the level 
densities are inputs in large computer codes and thousands of cross sections 
are estimated \cite{Go96}.

Also the present knowledge of the $\gamma$ strength function is poor. Although
the strengths can be roughly calculated by the Weisskopf estimate, which is
based on single particle transitions (see e.g.\ \cite{BM69}), some transitions
deviate many orders of magnitude from this approximation. A compilation of
average $\gamma$ transition strengths for dipole and electric quadrupole 
transitions can be found in \cite{ZS93}. The uncertainty of the $\gamma$ 
strength function concerns the absolute value and the $\gamma$ energy 
dependence. For E1 transitions one assumes that the $\gamma$ energy dependence
follows the Giant Dipole Resonance (GDR) $(\gamma,\gamma')$ cross section. This
is, however, to be proven.

In this work, we determine the level density and the $\gamma$ strength function
for $^{162}$Dy for energies close up to the neutron binding energy $B_n$. By 
comparing the present data, which were obtained from the 
$^{162}$Dy($^3$He,$^3$He'$\gamma$)$^{162}$Dy reaction, to previous data 
\cite{TB96,SB99b}, which were obtained from the 
$^{163}$Dy($^3$He,$\alpha\gamma$)$^{162}$Dy reaction, we can test if the basic
assumption of our analysis method is fulfilled.

This main assumption is that the $\gamma$ decay pattern from any excitation 
energy bin is independent of the population mechanism of states within this 
bin, e.g.\ direct population by a nuclear reaction, or indirect population by a
nuclear reaction followed by one or several $\gamma$ rays. Since the $\gamma$ 
decay probabilities of an excited state are independent of the populating 
reaction, the assumption above is generally equivalent to the assumption that 
the same states are populated equally by the direct and indirect population 
mechanisms. One can now imagine several cases where this assumption might be 
invalid. 

Firstly, thermalization time might compete with the half life of excited 
states, and the selectivity of the direct population by a nuclear reaction will
be reflected by a different $\gamma$ decay pattern with few and relatively 
strong $\gamma$ transitions compared to a statistical spectrum which is the
expected $\gamma$ decay pattern after complete thermalization. 

Secondly, direct population might populate states with different exact or
approximate quantum numbers like spin or parity than indirect population. Since
states with different exact or approximate quantum numbers do not mix at all or
very weakly in the latter case, the ensemble of populated states after 
thermalization will differ for the two population mechanisms and therefore one 
can expect different $\gamma$ decay patterns. 

It is very difficult to judge where the assumption of the method is applicable 
and how good this approximation is. Below, we will, by comparing two different 
direct population mechanisms represented by two different nuclear reactions, 
investigate in which excitation energy interval the assumption might break 
down. 

\section{Experiment and data analysis}

The experiment was carried out at the Oslo Cyclotron Laboratory (OCL) using the
MC35 Scanditronix cyclotron. The beam current was $\sim$1~nA of $^3$He 
particles with an energy of 45~MeV. The experiment was running for a total of 
2~weeks. The target was an isotopically enriched 95\% $^{162}$Dy self 
supporting metal foil with a thickness of 1.4~mg/cm$^2$ glued on an aluminum 
frame. Particle identification and energy measurements were performed by a ring
of 8~Si(Li) telescopes at 45$^\circ$ relative to the beam axis. The telescopes 
consist of a front and end detector with thicknesses of some 150~$\mu$m and 
3000~$\mu$m respectively, which is enough to effectively stop the ejectiles of 
the reaction. The $\gamma$ rays were detected by a ball of 27 5''$\times$5'' 
NaI(Tl) detectors (CACTUS) \cite{GA90} covering a solid angle of $\sim$15\% of 
$4\pi$. Three 60\% Ge(HP) detectors were used to monitor the selectivity of the
reaction and the entrance spin distribution of the product nucleus. During the 
experiment we collected besides data for the 
$^{162}$Dy($^3$He,$^3$He')$^{162}$Dy reaction, where results are presented in 
this work, also data for the $^{162}$Dy($^3$He,$\alpha$)$^{161}$Dy reaction, 
where some results were presented in \cite{SB99b,GB99}. A comprehensive 
description of the $^{163}$Dy($^3$He,$\alpha\gamma$)$^{162}$Dy experiment, 
which we will compare our findings to, can be found in \cite{GB94}. 

\begin{figure}[htbp]\centering
\mbox{\psfig{figure=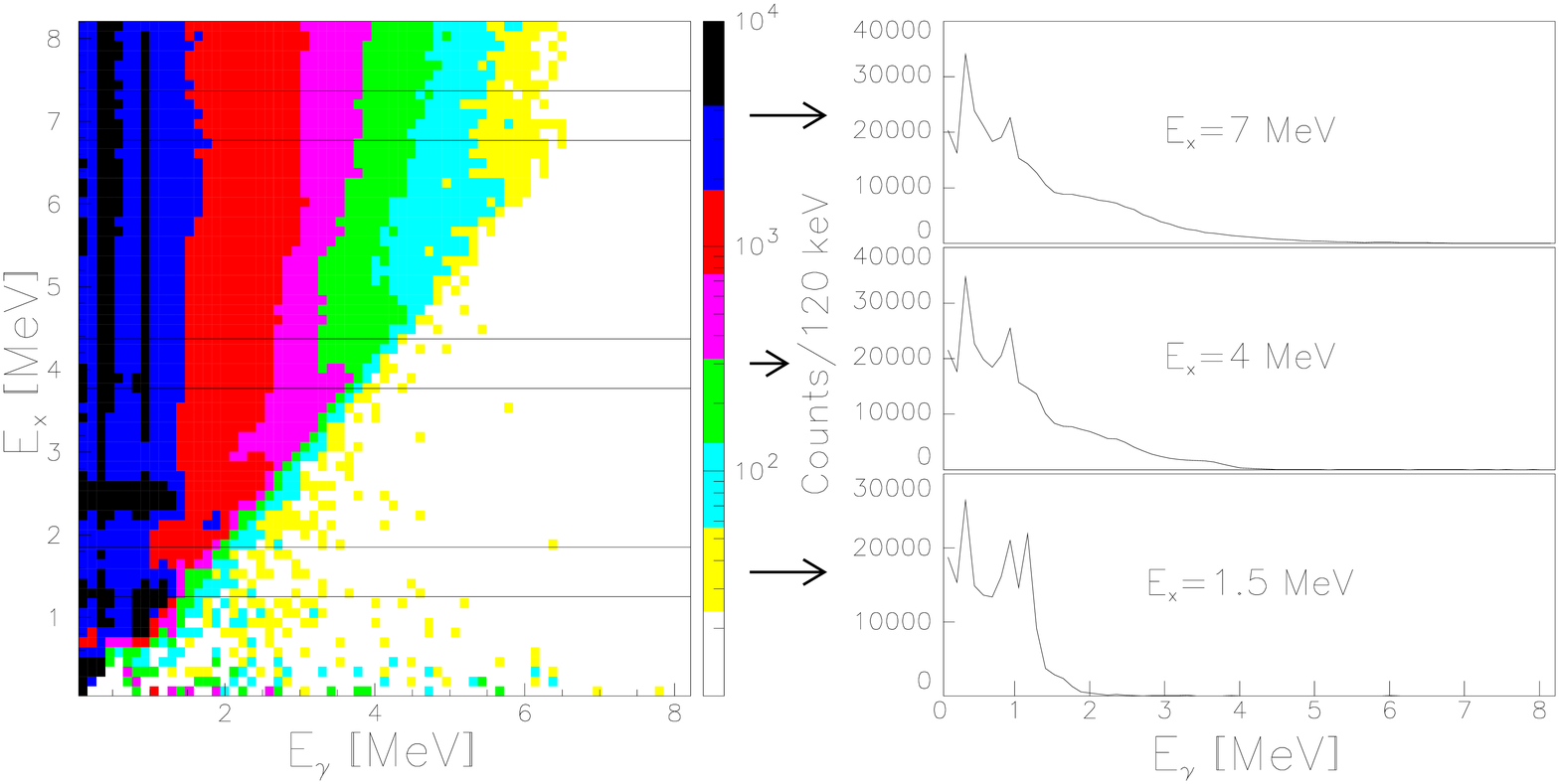,height=6cm}}
\caption{Raw data of the $^{162}$Dy($^3$He,$^3$He'$\gamma$)$^{162}$Dy reaction.
Some $\gamma$ spectra for different excitation energies are projected out on 
the right hand side.}
\label{fig:original}
\end{figure}

\begin{figure}[htbp]\centering
\mbox{\psfig{figure=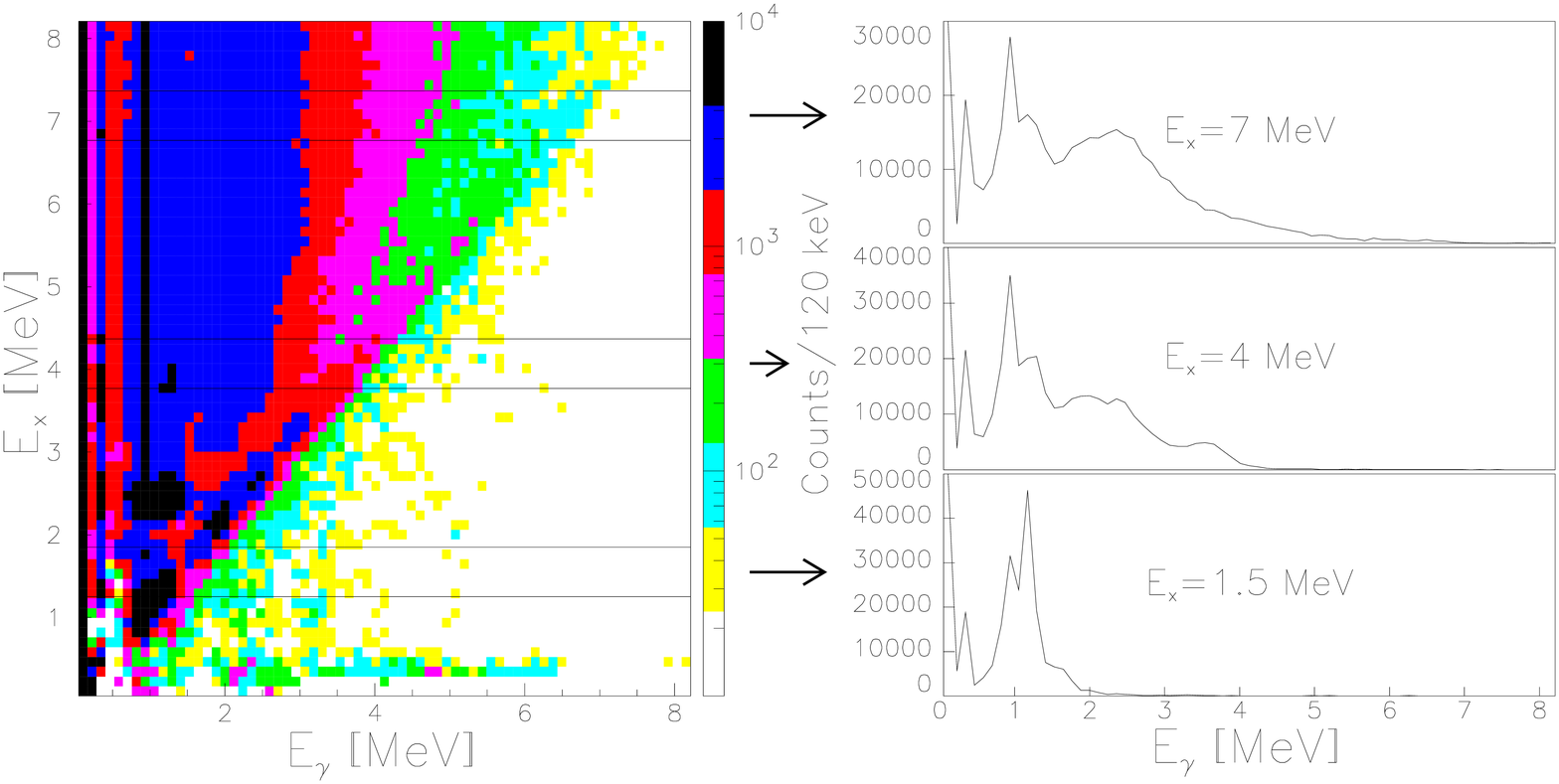,height=6cm}}
\caption{Unfolded data of the $^{162}$Dy($^3$He,$^3$He'$\gamma$)$^{162}$Dy
reaction. Also here, some $\gamma$ spectra for different excitation energies 
are projected out on the right hand side.}
\label{fig:unfolded}
\end{figure}

\begin{figure}[htbp]\centering
\mbox{\psfig{figure=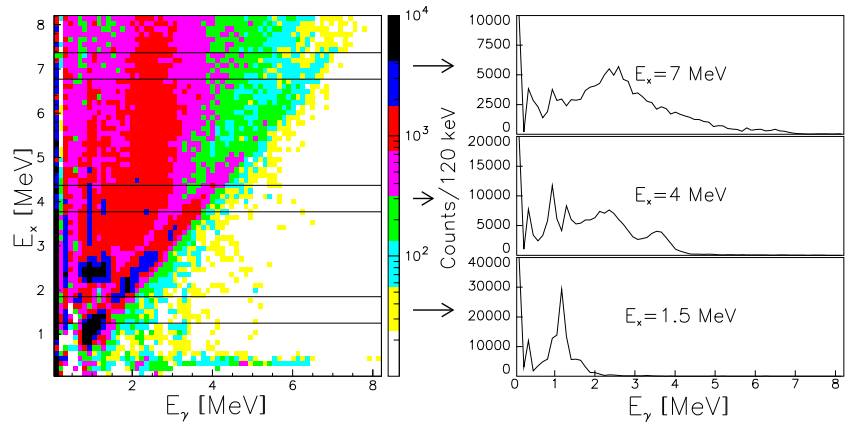,height=6cm}}
\caption{Primary $\gamma$ spectra of the 
$^{162}$Dy($^3$He,$^3$He'$\gamma$)$^{162}$Dy reaction. Here again, some 
$\gamma$ spectra for different excitation energies are projected out on the 
right hand side.}
\label{fig:fg}
\end{figure}

In the first step of the data analysis, the measured ejectile energy is 
transformed into excitation energy of the product nucleus. In 
Fig.~\ref{fig:original} the raw data are shown. In the next step, the $\gamma$ 
spectra are unfolded for every excitation energy bin using measured response 
functions of the CACTUS detector array \cite{GT96}. In Fig.~\ref{fig:unfolded} 
the unfolded data are shown. In the third step, the primary $\gamma$ spectra 
for every excitation energy bin are extracted from the unfolded data by the 
subtraction technique of Ref.~\cite{GR87}. In Fig.~\ref{fig:fg} the primary 
$\gamma$ spectra are shown.

\begin{figure}[htbp]\centering
\mbox{\psfig{figure=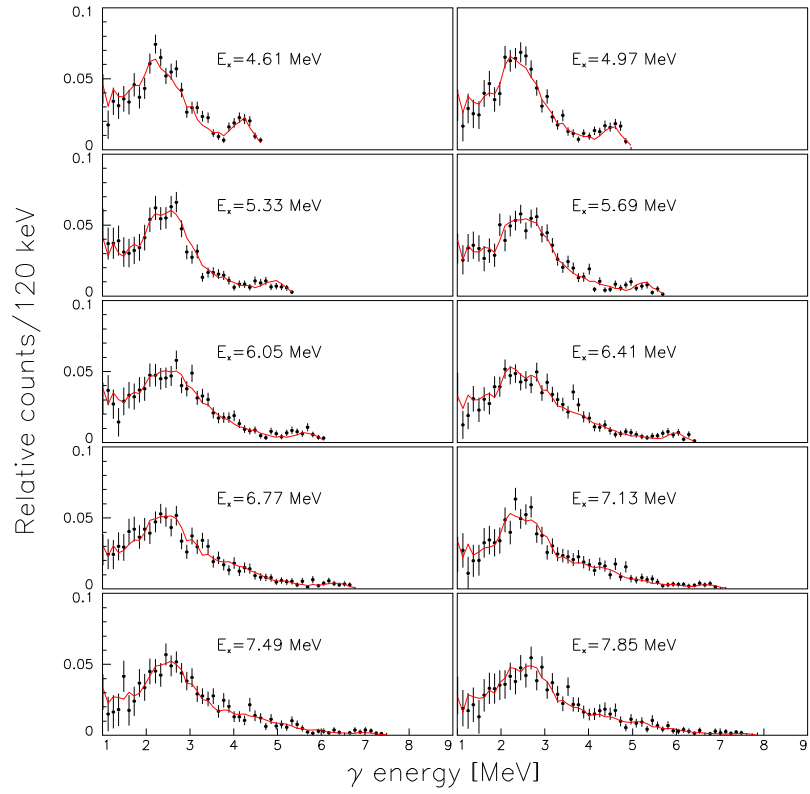,height=12.1cm}}
\caption{Normalized primary $\gamma$ spectra for the 
$^{162}$Dy($^3$He,$^3$He'$\gamma$)$^{162}$Dy reaction including estimated 
errors (data points) compared to the least $\chi^2$ fit according to 
Eq.~(\protect{\ref{eq:ansatz}}) (lines).}
\label{fig:fgboth}
\end{figure}

In the fourth step, we extract level density and $\gamma$ strength function 
from the primary $\gamma$ spectra. The main assumption behind this method is 
the Axel Brink hypothesis \cite{Br55,Ax62}
\begin{equation}
\label{eq:ansatz}
\Gamma(E_x,E_\gamma)\propto F(E_\gamma)\,\varrho(E_f)
\end{equation}
with $E_f=E_x-E_\gamma$. It says that the $\gamma$ decay probability in the 
continuum energy region represented by the primary $\gamma$ spectrum $\Gamma$ 
is proportional to the level density $\varrho$ and a $\gamma$ energy dependent 
factor $F$. The level density and the $\gamma$ energy dependent factor are 
estimated by a least $\chi^2$ fit to the experimental data \cite{SB99}. In 
Fig.~\ref{fig:fgboth} the experimental data including estimated errors 
\cite{SB99} are compared to the fit according to Eq.~(\ref{eq:ansatz}).

The data are fitted very well by the theoretical expression of 
Eq.~(\ref{eq:ansatz}). This is a remarkable example for the validity of the 
Axel Brink hypothesis. However, it can never completely be ruled out, that a 
minor portion of the primary $\gamma$ matrix cannot be factorized into a level 
density and a $\gamma$ energy dependent factor. One might also encounter large 
fluctuations in these quantities at very low level densities around the ground 
state or when considering highly collective $\gamma$ transitions and single 
particle $\gamma$ transitions at similar $\gamma$ energies.

Since the least $\chi^2$ fit according to Eq.~(\ref{eq:ansatz}) yields 
an infinitely large number of equally good solutions, which can be obtained by
transforming one arbitrary solution by
\begin{eqnarray}
\label{eq:trafo}
\tilde{\varrho}(E_x-E_\gamma)&=&\varrho(E_x-E_\gamma)\,
A\exp(\alpha[E_x-E_\gamma])\\
\tilde{F}(E_\gamma)&=&F(E_\gamma)\,B\exp(\alpha E_\gamma),
\end{eqnarray}
\cite{SB99} we have to determine the three parameters $A$, $B$ and $\alpha$ of 
the transformation by comparing the results to other experimental data. We fix 
the parameters $A$ and $\alpha$ by comparing the extracted level density curve 
to the number of known levels per excitation energy bin around the ground state
\cite{FS96} and to the level density at the neutron binding energy $B_n$ 
calculated from neutron resonance spacing data \cite{LH75}. Since the procedure
is described in detail in Ref.~\cite{SB99}, we only show in 
Fig.~\ref{fig:counting} how the extracted level density curve compares to other
experimental data.

The parameter $B$ could now in principle be fixed by comparing the extracted
$\gamma$ energy dependent factor $F$ to other experimental data of the $\gamma$
strength function. However since data are very sparse and the absolute 
normalization of $\gamma$ strength function data is very uncertain, we give the
$\gamma$ energy dependent factor in arbitrary units.

\section{Results and discussion}

\subsection{The level density}

\begin{figure}[htbp]\centering
\mbox{\psfig{figure=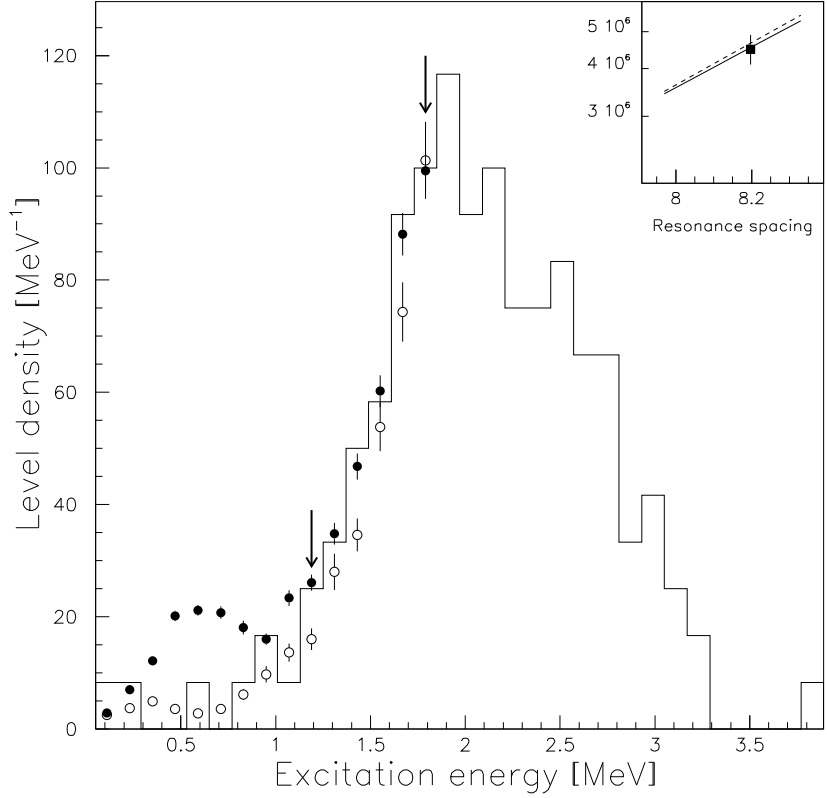,height=12.1cm}}
\caption{Determination of parameters $A$ and $\alpha$ of 
Eq.~(\protect{\ref{eq:trafo}}). The extracted level density curve from the 
$^{162}$Dy($^3$He,$^3$He'$\gamma$)$^{162}$Dy reaction data (full data points 
and line in insert) is compared to the number of known levels per excitation 
energy bin around the ground state (histogram) in the region between the 
arrows, and to the level density at the neutron binding energy $B_n$, 
calculated from neutron resonance spacing data (square in insert). In 
comparison, the extracted level density curve from the 
$^{163}$Dy($^3$He,$\alpha\gamma$)$^{162}$Dy reaction data (empty data points
and slashed line in insert) is shown.}
\label{fig:counting}
\end{figure}

\begin{figure}[htbp]\centering
\mbox{\psfig{figure=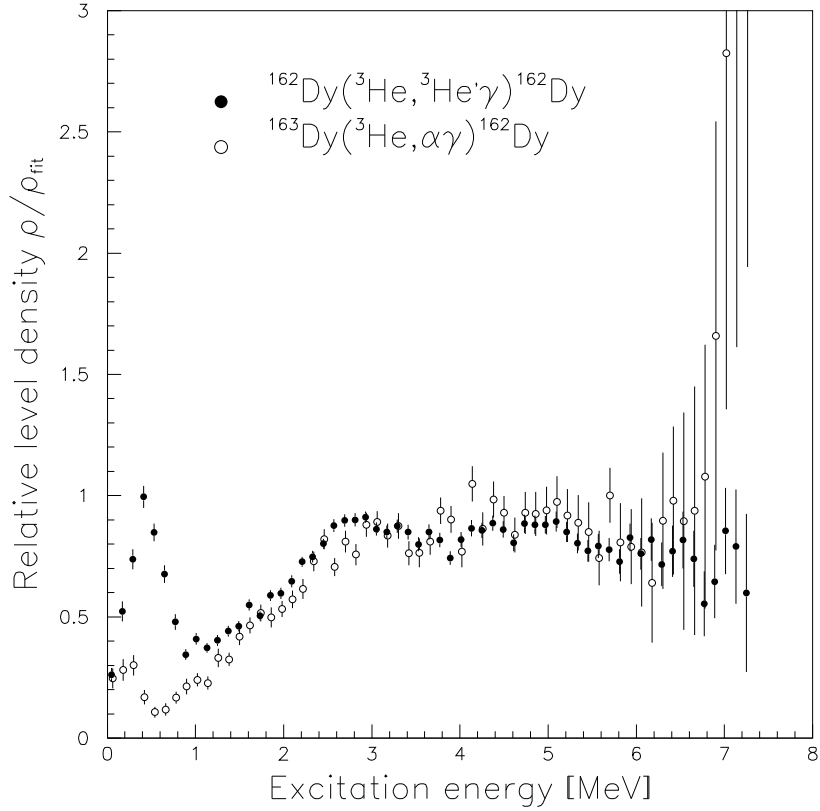,height=12.1cm}}
\caption{Comparison of the extracted relative level density of $^{162}$Dy 
deduced from the $^{162}$Dy($^3$He,$^3$He'$\gamma$)$^{162}$Dy reaction (this 
work) and from the $^{163}$Dy($^3$He,$\alpha\gamma$)$^{162}$Dy reaction 
(previous works). The error bars of the former level density curve are about 
half of the errors of the latter due to $\sim$5 times better statistics in the 
data of the $^{162}$Dy($^3$He,$^3$He'$\gamma$)$^{162}$Dy reaction. The 
differences between the two curves below 1.5~MeV of excitation energy are 
accounted for in the text.}
\label{fig:rho}
\end{figure}

We compare extracted level densities of $^{162}$Dy from two reactions, namely 
$^{162}$Dy($^3$He,$^3$He'$\gamma$)$^{162}$Dy and 
$^{163}$Dy($^3$He,$\alpha\gamma$)$^{162}$Dy. While level densities from the 
latter reaction were already published in \cite{HB95,TB96,MB99} using 
approximate extraction methods, and in \cite{SB99b} in the present form, data 
from the first reaction are shown here for the first time. Figure~\ref{fig:rho}
shows the relative level densities, which are calculated by dividing the 
extracted level densities by an exponential $C\exp(E/T)$ with $T=580$~keV and 
$C=10$~MeV$^{-1}$ in our case. One can see that both level densities agree very
well within 10\% in the excitation energy interval 1.5~MeV to 6.5~MeV. This 
result is very encouraging, since level densities are generally only known 
within an error of $\pm$50-100\%. Above 6.5~MeV the errors are too large in 
order to make conclusive observations. Below $\sim$1.5~MeV the two level 
densities differ dramatically from each other. In Fig.~\ref{fig:counting} 
one can see that the extracted level density from the 
$^{163}$Dy($^3$He,$\alpha\gamma$)$^{162}$Dy reaction agrees very well with the 
number of known levels per excitation energy bin below $\sim$1.2~MeV, whereas 
the extracted level density from the 
$^{162}$Dy($^3$He,$^3$He'$\gamma$)$^{162}$Dy reaction overestimates the number 
of levels in this energy region by a factor of $\sim$3.

The level density at $\sim$0.5~MeV of excitation energy is determined by the
data in the primary $\gamma$ matrix which lie approximately on the diagonal
$E_x\stackrel{>}{\sim}E_\gamma$ (see Fig.~\ref{fig:fg}). Careful examination of
Fig.~\ref{fig:fgboth} shows, that the bumps at $E_x\stackrel{>}{\sim}E_\gamma$
are very well fitted by the factorization given by Eq.~(\ref{eq:ansatz}). We 
therefore conclude, that the differences in level density around $\sim$0.5~MeV
of excitation energy are not artifacts of the extraction method, but have their
origin in differences of the primary $\gamma$ spectra. We actually find in the 
primary $\gamma$ matrix of the $^{162}$Dy($^3$He,$^3$He'$\gamma$)$^{162}$Dy 
reaction a large number of high energetic $\gamma$ transitions, connecting the 
direct populated states with the ground state rotational band. This surplus of 
counts compared to primary $\gamma$ spectra from the 
$^{163}$Dy($^3$He,$\alpha\gamma$)$^{162}$Dy reaction is the reason for 
overestimating the level density at $\sim$0.5~MeV of excitation energy. 

We argue that the level density curve extracted from the neutron pick up 
reaction data is the more realistic one, as supported by 
Fig.~\ref{fig:counting}. Since the neutron pick up reaction cross section is 
dominated by high $l$ neutron transfer, the direct population of the $^{162}$Dy
nucleus takes place through one particle one hole components of the wave 
functions. Such configurations are not eigenstates of the nucleus, but they are
rather distributed over virtually all eigenstates in the neighboring 
excitation energy region. Thus, we can expect fast and complete thermalization 
before $\gamma$ emission. The inelastic $^3$He scattering on the other hand is 
known to populate mainly collective excitations. These collective excitations 
will thermalize rather slowly, since their structure is much more like 
eigenstates of the nucleus, and their wave functions are less spread over 
eigenfunctions in the close excitation energy region. However, we can expect 
that their structure is similar to the structure of states in the ground state 
rotational band. Therefore, the large $\gamma$ transition rates from the direct
populated states to the ground state rotational band might just reflect the 
inverse process of inelastic scattering. The surplus of $\gamma$ counts can 
therefore be interpreted as preequilibrium decay. An extreme example for this 
are nuclear resonance fluorescence studies (NRF) \cite{WB88}. It is estimated, 
that in even even nuclei more than 90\% of the $\gamma$ strength from states 
excited by $\gamma$ rays is going to the ground state or to the first excited 
state. Thermalization of the excited states in NRF is also hindered by the fact
that one populates isovector states, which in the proton neutron interacting
boson model (IBA-2) are characterized by a different (approximate) $F$ spin 
quantum number than other states in the same excitation energy regions.  

We would like to point out, that although the basic assumption behind the
primary $\gamma$ method is partially violated in the case of the 
$^{162}$Dy($^3$He,$^3$He'$\gamma$)$^{162}$Dy reaction, the level densities in 
the excitation energy interval 1.5~MeV to 6.5~MeV deduced from the two 
reactions agree extremely well. This indicates, that the extracted level 
density curves are quite robust with respect to the goodness of the assumption.
Especially the bump at $\sim$2.5~MeV excitation energy indicating the breaking 
of nucleon pairs \cite{MB99,GB99} and the quenching of pairing correlations 
\cite{SB99b} could be very well reproduced. One should also keep in mind that
the two reactions populate states with slightly different spin distributions
due to the different target spins in the two reactions, which might account for
some differences in the extracted level densities.

\subsection{The $\gamma$ energy dependent factor}

\begin{figure}[htbp]\centering
\mbox{\psfig{figure=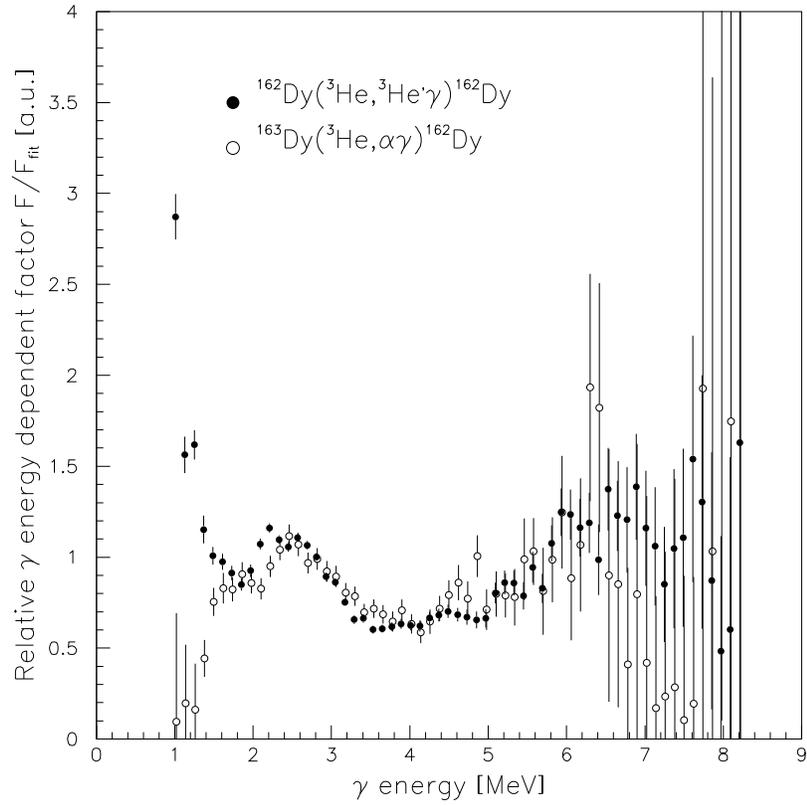,height=12.1cm}}
\caption{Comparison of the extracted relative $\gamma$ energy dependent 
function of $^{162}$Dy deduced from the 
$^{162}$Dy($^3$He,$^3$He'$\gamma$)$^{162}$Dy (this work) and from the
$^{163}$Dy($^3$He,$\alpha\gamma$)$^{162}$Dy reaction (previous work, reanalyzed
in this work). Also here, the error bars of the relative $\gamma$ energy 
dependent function extracted from the data of the former reaction are about 
half of the other ones, due to better statistics.} 
\label{fig:sig}
\end{figure}

We compare the extracted $\gamma$ energy dependent function $F$ of $^{162}$Dy
for the two reactions. The $F$ function from the 
$^{163}$Dy($^3$He,$\alpha\gamma$)$^{162}$Dy reaction was already published in 
\cite{TB96} using an approximate extraction method, however the data were 
reanalyzed using the exact extraction method of Ref.~\cite{SB99} and are in the
present form, as well as data from the 
$^{162}$Dy($^3$He,$^3$He'$\gamma$)$^{162}$Dy reaction, published for the first 
time in this work. Figure~\ref{fig:sig} shows the relative $F$ functions, which
are obtained by dividing the extracted $F$ function by $E_\gamma^n$ with 
$n=4.3$ and scaling them to $\sim$1 at $\sim$4~MeV of $\gamma$ energy. Also in 
this case the two functions agree within 10\% in the $\gamma$ energy interval
of 1.5~MeV to 6.5~MeV. Above $\sim$6.5~MeV again, the error bars are too large 
in order to allow for any conclusions. Below $\sim$1.3~MeV of $\gamma$ energy, 
the two functions differ dramatically from each other. Due to experimental 
difficulties, like ADC threshold walk and bad timing properties of low 
energetic $\gamma$ rays, we had to exclude $\gamma$ rays with energies below 
1~MeV from the data analysis \cite{SB99}. It is therefore very difficult to 
judge if the differences in the $F$ function curves below 1.5~MeV of $\gamma$ 
energy are also due to experimental problems (i.e.\ the experimental cut was 
too optimistic, and we should rather have excluded all $\gamma$ rays with 
energies below 1.5~MeV) or due to the different nuclear reactions used to 
excite the $^{162}$Dy nucleus. 

Also here we would like to emphasize, that despite the basic assumption behind 
the primary $\gamma$ method is not completely fulfilled in the case of the
$^{162}$Dy($^3$He,$^3$He'$\gamma$)$^{162}$Dy reaction, the two $F$ functions
agree very well. Especially the bump at $\sim$2.5~MeV of $\gamma$ energy, which
we interpret as a Pigmy Resonance is equally pronounced in both reactions. We 
are therefore very confident that the extracted level density and $\gamma$ 
energy dependent factor for $^{162}$Dy presented in this work are not, or very 
little, reaction dependent.

\section{Conclusions}

This work compares the results from the 
$^{162}$Dy($^3$He,$^3$He'$\gamma$)$^{162}$Dy reaction to those of the 
$^{163}$Dy($^3$He,$\alpha\gamma$)$^{162}$Dy reaction. The level density 
$\varrho$ and the $\gamma$ energy dependent factor $F$ in $^{162}$Dy are shown
to be reliably extracted with our method in the energy interval 1.5-6.5~MeV. 
The findings are independent of the particular reaction chosen to excite the 
$^{162}$Dy nucleus. The two reactions differ from each other (i) in the 
reaction type; i.e.\ inelastic $^3$He scattering versus neutron pick up, and 
thus in the nuclear states populated before thermalization, namely collective 
excitations versus one particle one hole states, (ii) in the target spins; 
$0^+$ for $^{162}$Dy versus $5/2^-$ for $^{163}$Dy, and thus in the spin 
distribution of direct populated states, and (iii) in the $Q$-value; 0~MeV for 
inelastic $^3$He scattering versus 14.3~MeV for the neutron pick up reaction. 
Nevertheless, the only differences in the extracted quantities are those in the
level densities below $\sim$1.5~MeV of excitation energy. These might be 
explained by preequilibrium $\gamma$ decay in the 
$^{162}$Dy($^3$He,$^3$He'$\gamma$)$^{162}$Dy reaction, whereas the 
$^{163}$Dy($^3$He,$\alpha\gamma$)$^{162}$Dy reaction is supposed to show only 
equilibrium $\gamma$ decay, and thus reveals reliable level densities below 
1.5~MeV of excitation energy, which is supported by comparison to known data. 
However, although preequilibrium $\gamma$ decay violates the basic assumption 
of the primary $\gamma$ method, the effect on the extracted level density 
$\varrho$ and the $\gamma$ energy dependent factor $F$ between 1.5~MeV and 
6.5~MeV of energy is shown to be less than 10\%. In conclusion, the present 
results have given further confidence in the new extraction techniques, and 
open for several interesting applications in the future.

The preequilibrium decay does not seem to violate the Axel Brink hypothesis, 
since the respective parts of the primary $\gamma$ spectrum could be fitted 
within this assumption. However, the extracted quantities $\varrho$ and $F$ 
will then only represent a weighted sum of the respective quantities obtained 
from preequilibrium and equilibrium $\gamma$ decay, where in the case of the 
$^{162}$Dy($^3$He,$^3$He'$\gamma$)$^{162}$Dy reaction, the preequilibrium 
process dominates the level density below 1.5~MeV of excitation energy.
We conclude therefore that neutron pick up reactions are more suitable than 
inelastic $^3$He scattering for our method, since the states populated by the 
former reaction presumably thermalize completely, whereas those populated by 
the latter reaction might not completely thermalize before $\gamma$ emission.

\section{Acknowledgments}

The authors wish to thank Jette S{\"o}rensen for making the target and 
E.A.~Olsen and J.~Wikne for excellent experimental conditions. Financial 
support from the Norwegian Research Council (NFR) is gratefully acknowledged.

\end{document}